\documentclass[reprint,showpacs]{revtex4-1}

\usepackage[pdftex]{graphicx}

\begin{document}

\title{Self-normalizing phase measurement in multimode terahertz spectroscopy based on photomixing of three lasers}

\author{K. Thirunavukkuarasu$^{1,2}$}
\author{M. Langenbach$^{1}$}
\author{A. Roggenbuck$^{1,3}$}
\author{E. Vidal$^{1}$}
\author{H. Schmitz$^{1}$}
\author{J.~Hemberger$^{1}$}
\author{M.~Gr{\"u}ninger$^{1}$}
\affiliation{$^1$ II. Physikalisches Institut,
Universit\"{a}t zu K\"{o}ln, Z\"{u}lpicher Stra{\ss}e 77, D-50937 K\"{o}ln, Germany}
\affiliation{$^2$ National High Magnetic Field Laboratory, Tallahassee, Florida 32310, USA}
\affiliation{$^3$ TOPTICA Photonics AG, Lochhamer Schlag 19, D-82166 Gr\"{a}felfing, Germany}

\begin{abstract}
Photomixing of two near-infrared lasers is well established for continuous-wave terahertz spectroscopy.
Photomixing of three lasers allows us to measure at three terahertz frequencies simultaneously.
Similar to Fourier spectroscopy, the spectral information is contained in an interferogram, which is equivalent to
the waveform in time-domain spectroscopy.
We use one fixed terahertz frequency $\nu_{\rm ref}$ to monitor temporal drifts of the setup, i.e.,
of the optical path-length difference.
The other two frequencies are scanned for broadband high-resolution spectroscopy.
The frequency dependence of the phase is obtained with high accuracy by normalizing it to the
data obtained at $\nu_{\rm ref}$, which eliminates drifts of the optical path-length difference.
We achieve an accuracy of about $1-2\,\mu$m or $10^{-8}$ of the optical path length.
This method is particularly suitable for applications in nonideal environmental conditions
outside of an air-conditioned laboratory.
\end{abstract}


\date{October 2, 2014}
\maketitle

Terahertz spectroscopy has been revolutionized by laser-based techniques and bears an enormous potential
both for fundamental science and for a wide range of applications.\cite{Lee08,Tonouchi2007}
One intriguing aspect of the terahertz range is that it allows for the determination of both amplitude
and phase $\varphi$ of an electromagnetic wave.
The phase delay induced by a sample can be employed for measuring both the refractive index and the
thickness\cite{Jen2014,Scheller2010a,Yasui2005,Johnson2001,Roggenbuck2010}
or for tomography, e.g.\ for the inspection of space shuttle foam insulation.\cite{Mittleman1997,Zhong2005,Guillet2014}
Reliable measurements of the phase require a high stability of the relevant experimental lengths,
therefore (thermal) fluctuations or alignment drifts may not exceed a small fraction of the wavelength.
Large wavelengths such as about 1\,mm at 300\,GHz thus facilitate the determination of the phase.
However, thermal fluctuations cannot be fully suppressed even in an air-conditioned laboratory,
and phase measurements in real-world applications in a less than ideal environment are challenging,
in particular if they rely on a robust fiber-based system.\cite{Soltani2014}
These difficulties are successfully surpassed in ellipsometry, which measures the phase \textit{difference}
between different polarization states.
Here, we choose another route based on continuous-wave (cw) spectroscopy in the frequency domain
and consider the phase difference of waves with different frequencies.
Via photomixing of three lasers, we generate waves at three terahertz frequencies.
The waves travel along \textit{the same path at the same time}, their phases are measured simultaneously.
We employ the phase at the fixed frequency $\nu_{\rm ref}$ to monitor length changes during the measurement.
The normalized phases of the two other, scanning frequencies are nearly insensitive to thermal drifts.
Without temperature stabilization of the laboratory, we achieve an accuracy which is equivalent to length changes
of about 1-2\,$\mu$m or 3-6\,fs$\cdot c$, where $c$ denotes the speed of light.

\begin{figure}
\centering
\includegraphics[width=0.95\columnwidth]{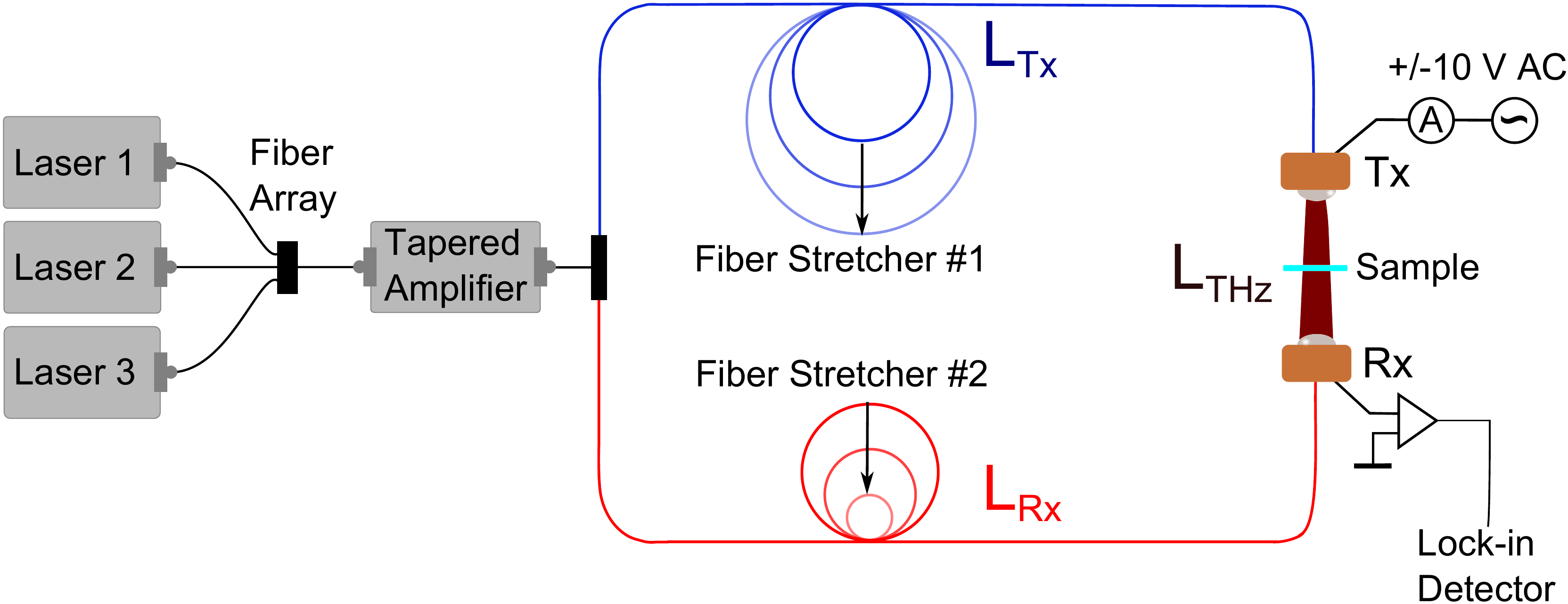}
\caption{Sketch of the setup with three lasers.
The fiber array feeds 1/3 of the power of each laser into the amplifier.
A face-to-face setup of the photomixers yields a short terahertz path length $L_{\rm THz}$.
The optical path-length difference is modulated by two fiber stretchers.
}
\label{fig:3LaserSetup}
\end{figure}

Continuous-wave terahertz radiation can be generated and coherently detected
by illuminating two photomixers, transmitter and receiver,
with the optical beat of two near-infrared lasers with frequencies $\nu_1$
and $\nu_2$.\cite{McIntosh1995,Verghese1998,Matsuura2005,Deninger2008,Roggenbuck2010}
The biased transmitter emits radiation at the difference frequency $\nu$\,=\,$|\nu_2-\nu_1|$,
whereas the photocurrent $I_\mathrm{ph}$ in the receiver measures the cross-correlation\cite{Verghese1998,Morikawa2000}
\begin{equation}
    I_\mathrm{ph}\propto E_\mathrm{THz}\cos(\Delta\varphi) \, ,
\label{eq:Iph}
\end{equation}
where $\Delta \varphi$ denotes the phase difference between the optical beat and the terahertz electric field
with amplitude $E_\mathrm{THz}$ at the receiver.
The sensitivity to $\cos(\Delta\varphi)$ reflects the similarity to a Mach-Zehnder interferometer.
Both $E_{\rm THz}$ and $\Delta\varphi$ can be determined from $I_{\rm ph}$ by phase modulation with, e.g.,
a mechanical delay stage or a fiber stretcher.\cite{Roggenbuck2012}
In both cases, one modulates the optical path-length difference
$\Delta L$\,=\,$L_{\rm Tx}+L_{\rm THz}-L_{\rm Rx}$
between the receiver arm with the optical path length $L_{\rm R_x}$ and the transmitter arm including the terahertz path,
$L_{\rm Tx} + L_{\rm THz}$, see Fig.\ \ref{fig:3LaserSetup}.
A frequency-independent $\Delta L$ contributes a term $\propto \nu$ to $\Delta \varphi$,
but there are other contributions
stemming from, e.g., the group delay introduced by the antennae, the photomixer impedance,
and standing waves. \cite{Langenbach2014} We summarize these in $\Delta \varphi_0(\nu)$,
\begin{equation}
  \Delta\varphi(\nu) = \Delta\varphi_0(\nu) + \Delta L \cdot \frac{2\pi \nu}{c} \, .
\label{eq:DelPhi}
\end{equation}
Accordingly, an uncertainty $\delta \varphi$ of $\Delta \varphi$ arises from
the uncertainties $\delta L$ and $\delta \nu$ of $\Delta L$ and $\nu$, respectively,
\begin{equation}
  \delta\varphi \cdot \frac{c}{2\pi\, \nu} =  \delta L
  + \left[ \Delta L + \frac{\partial \Delta \varphi_0}{\partial \nu} \frac{c}{2\pi}\right] \frac{\delta\nu}{\nu} \, .
\label{eq:delPhi}
\end{equation}
In our setup, the right hand side typically is dominated by the first term, $\delta L$.\cite{Langenbach2014}
The line width of the beat signal of two tunable lasers amounts to about $\delta \nu$\,=\,5\,MHz.\cite{Deninger2008}
Choosing, e.g., $|\Delta L|$\,$\leq$\,1\,cm and $L_{\rm THz} \! \approx \! 20$\,cm,
the second term on the right hand side roughly yields 0.12\,$\mu$m$\cdot$(THz/$\nu$)$^2$, which amounts to
about 3\,$\mu$m at 200\,GHz and 0.75\,$\mu$m at 400\,GHz.\cite{Langenbach2014}
Typically, one has to cope with much larger values of the length drift $\delta L$,
which thus dominates $\delta \varphi/\nu$.
This claim is justified \textit{a posteriori} by the success of our normalization procedure.

\begin{figure}
\center
\includegraphics[width=0.5\columnwidth]{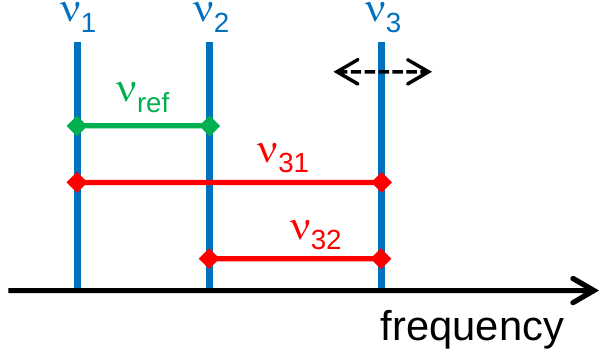}
\caption{Sketch of the three near-infrared laser frequencies $\nu_1$, $\nu_2$, and $\nu_3$ as well as of the
resulting difference frequencies in the terahertz range.
While $\nu_1$ and $\nu_2$ are fixed, $\nu_3$ can be scanned over a range of about 1\,THz. }
\label{fig:beatscheme}
\end{figure}

Continuous-wave terahertz spectroscopy based on photomixing of more than two modes has been discussed
previously.\cite{Morikawa2000,Tani2005,Gregory2005,Scheller2009,Brenner2010,Molter2011,Morikawa2011,Morikawa2013,Scheller2010a,Scheller2010b}
Using multimode laser diodes has been proposed as a low-cost, fast alternative for broadband
spectroscopy\cite{Morikawa2000,Tani2005,Gregory2005,Scheller2009,Brenner2010,Molter2011,Morikawa2011,Morikawa2013}
or to overcome the $2\pi$ ambiguity of the phase.\cite{Scheller2010a,Scheller2010b}
Here, we use \textit{three} near-infrared lasers to correct the length drift $\delta L$ in broadband spectroscopy.
Two lasers operate at the fixed frequencies $\nu_1$ and $\nu_2$
while the frequency $\nu_3$ of the third laser is tunable.
In the terahertz range, this yields three difference frequencies.\cite{Tani2005,Gregory2005,Scheller2009}
The reference frequency $\nu_{\rm ref}$\,=\,$|\nu_2 - \nu_1|$ is fixed
while $\nu_{31}$\,=\,$|\nu_3 - \nu_1|$ and $\nu_{32}$\,=\,$|\nu_3 - \nu_2|$
are tunable, see Fig.\ \ref{fig:beatscheme}.
Spectroscopic measurements are performed by varying $\nu_3$, thus tuning $\nu_{31}$ and $\nu_{32}$
over the desired range. Simultaneously, we record the temporal evolution of
$\delta \varphi (\nu_{\rm ref})$.
This yields an excellent measure of the drift $\delta L$
which allows for an accurate correction of the spectroscopic data.

A sketch of our experimental setup is given in Fig.\ \ref{fig:3LaserSetup}.
We employ three distributed-feedback diode lasers (\textsc{Toptica} DL DFB) with center wavelengths of about 780\,nm.
One laser is locked to a Doppler-free Rb absorption line at $\nu_1$\,$\approx$\,$c$/780\,nm with a stability of about 1\,MHz.
The other two lasers with frequencies $\nu_2$ and $\nu_3$ are tunable over a broad range,
the difference frequencies $|\nu_3 - \nu_1|$ and $|\nu_3 - \nu_2|$ cover the range up to 1.8\,THz.
Although our method is based on two fixed lasers at $\nu_1$ and $\nu_2$ plus a tunable one at $\nu_3$,
it is advantageous if both $\nu_2$ and $\nu_3$ are tunable.
In this way the reference frequency $\nu_{\rm ref}$\,=\,$|\nu_2 - \nu_1|$ can be adapted to the
experimental conditions, e.g., the transparent frequency range of a given sample.
First, we tune $\nu_2$ to choose an appropriate value of $\nu_{\rm ref}$.
Then, spectroscopic measurements are performed by scanning $\nu_3$ over a wide range.

The laser beams are superimposed in a polarization-maintaining single-mode fiber array.
Two beams are coupled in a 50:50 splitter, the third one is added via a 2:1 splitter.
The superposition thus carries 1/3 of the power of each laser.
After amplification (\textsc{Toptica} BoosTA 780), we use a further 50:50 splitter to illuminate
the two fiber-pigtailed photomixers.\cite{Mayorga2007}
For more details of the two-laser setup, we refer to
Refs.\ \onlinecite{Roggenbuck2010,Roggenbuck2012,Roggenbuck2013,Langenbach2014}.

\begin{figure}
\centering
\includegraphics[width=0.7\columnwidth]{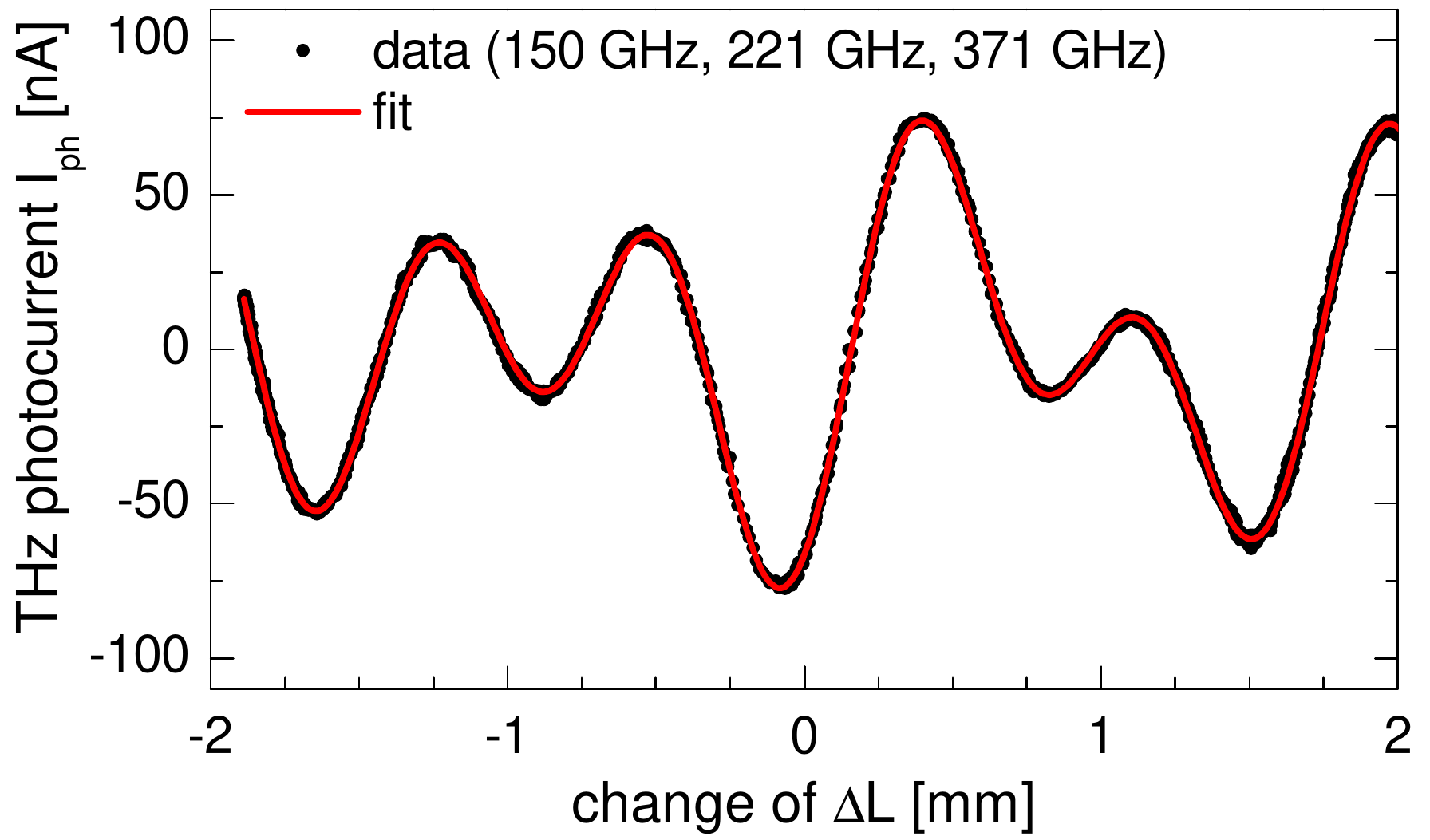}
\caption{The interferogram or photocurrent $I_{\rm ph}(\Delta L)$ is equivalent to the waveform
in a time-domain experiment.
Black: data for a representative choice of frequencies. Red: fit.
}
\label{fig:stretcherdata}
\end{figure}

A temporal modulation of $\Delta L$ yields a photocurrent $I_{\rm ph}(\Delta L)$ which is equivalent to an interferogram.\cite{Verghese1998}
In the conventional case of a two-laser setup, there is a single difference frequency and
the interferogram takes a cosine shape, see Equs.\ \ref{eq:Iph} and \ref{eq:DelPhi}.
For three difference frequencies, $I_{\rm ph}(\Delta L)$ is a superposition of
three cosines,\cite{Tani2005,Gregory2005,Scheller2009} see the example in Fig.\ \ref{fig:stretcherdata}.
We modulate $\Delta L$
by two fiber stretchers which affect $L_{\rm Rx}$ and $L_{\rm Tx}$ with opposite signs.\cite{Roggenbuck2012}
Each fiber stretcher adds 60\,m of fiber with a refractive index of about 1.5 to the optical path,
thus a single stretcher would introduce a large length difference between $L_{\rm Rx}$ and $L_{\rm Tx}$.
Using two fiber stretchers allows for a small value of $\Delta L$ and minimizes thermal drifts of $\Delta L$.
We employ a modulation frequency of 800\,Hz and a delay length or maximum change of $\Delta L$ of
$D \! \approx \! 5$\,mm.\cite{Roggenbuck2012}
Typically, we collect data for a single interferogram over about 240 stretcher cycles (or 300\,ms),
resulting in a net data acquisition rate of about 3\,Hz.\cite{Roggenbuck2012}

In the photocurrent $I_{\rm ph}(\Delta L)$, the cosine amplitudes depend on the respective frequencies
due to, e.g., the frequency dependence of the photomixers.\cite{Deninger2008,Roggenbuck2010}
A Fourier transform or a fit of the interferogram yields the amplitudes and phases for all three difference frequencies.
The spectral resolution d$\nu$\,=\,$c/2D$ of an interferogram amounts to 30\,GHz for a delay length $D$\,=\,5\,mm.
In our case, the spectral resolution is given by the
line width of the beat signal, $\delta \nu$\,=\,5\,MHz, or by the long-term frequency stability
of better than 20\,MHz over 24\, h.\cite{Deninger2008}
A reliable fit of the interferogram still requires that the three terahertz frequencies differ by at least d$\nu$.
Therefore, spectroscopic data cannot be obtained in the window $\nu_{\rm ref} \pm {\rm d}\nu$.
We typically collect data sets for different values of $\nu_{\rm ref}$,
which is another reason for using a tunable laser frequency $\nu_2$, i.e.,
a tunable $\nu_{\rm ref}$\,=\,$|\nu_2 - \nu_1|$.

\begin{figure}
\centering
\includegraphics[width=0.8\columnwidth]{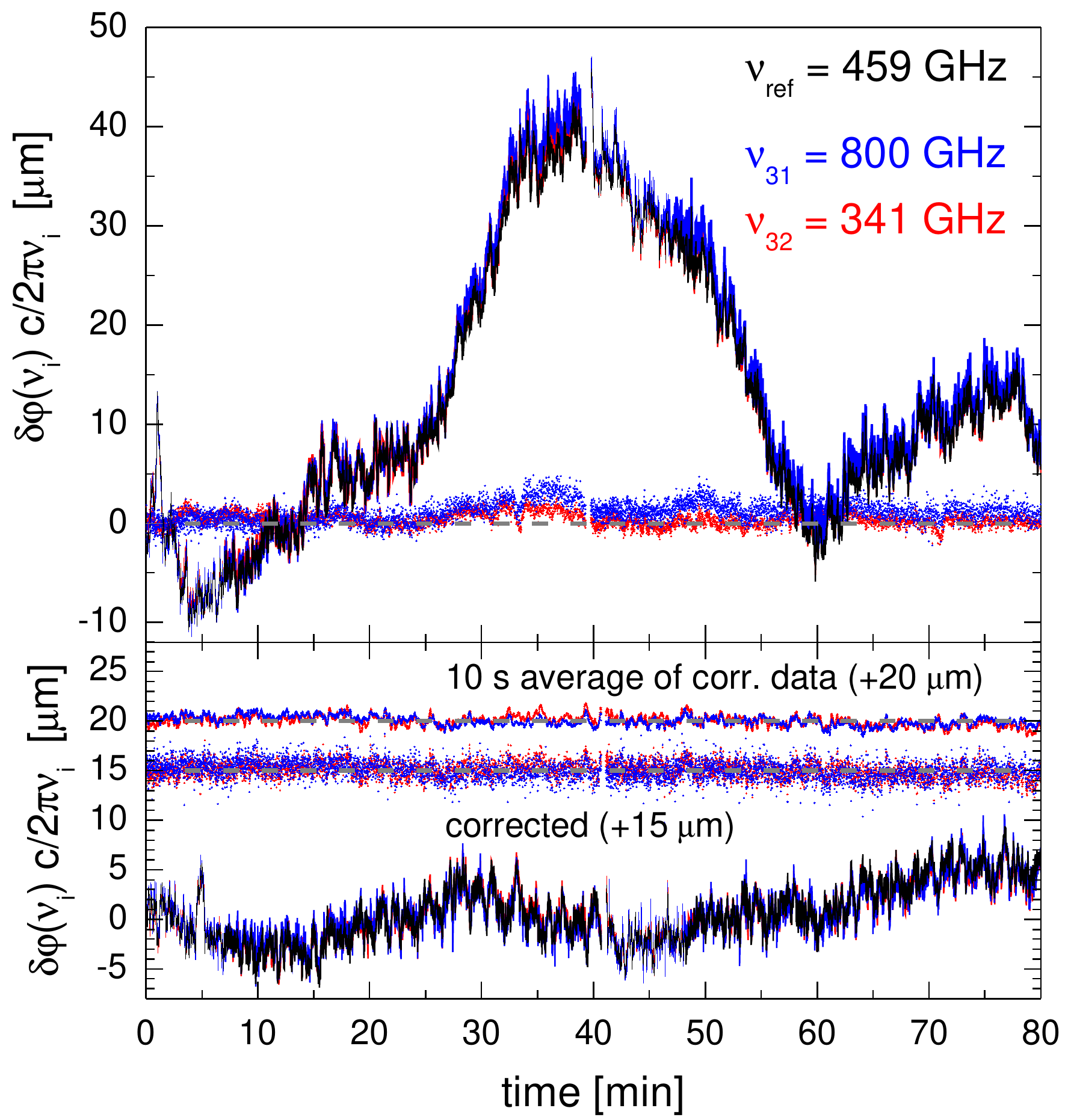}
\caption{Solid lines: Temporal drift $\delta L(t)$ of the optical path-length difference,
or, more precisely, of $\gamma_i(t)$ (cf.\ Eq.\ \ref{eq:delL}) measured with three constant terahertz frequencies $\nu_i$
for $|\Delta L| \! \leq \! 1$\,cm.
Symbols: corrected values $\gamma_{31}(t)-\gamma_{\rm ref}(t)$ (blue) and $\gamma_{32}(t)-\gamma_{\rm ref}(t)$ (red).
Data in top (bottom) panel were measured without (with) good thermal contact between the two fiber stretchers.
}
\label{fig:3freqcorr}
\end{figure}

Before addressing our results, we discuss the expected size of the drift.
First we consider a free-space setup without fibers. For $\Delta L$\,=\,0,
a common change $\delta T$ of the laboratory temperature equally affects
both interferometer arms, leaving  $\Delta L$ unchanged.
However, a temperature difference $\Delta T$ between the two arms induces a path-length difference of,
e.g., 20\,$\mu$m$\cdot(L_{\rm Rx}$/1m)$\cdot(\Delta T$/1K) if the setup is mounted
on an Al plate with a thermal expansion coefficient $\alpha_{\rm Al} \! \approx \! 2\cdot 10^{-5}$/K.\@
With dispersive elements such as the fibers, the discussion is more subtle.
To achieve $\Delta L \! \approx \! 0$, the free-space path length $L_{\rm THz}$ has to be compensated
by a piece of fiber, for which the thermal drift is dominated\cite{Roggenbuck2012} by the thermo-optic coefficient
$\frac{\partial n}{\partial T} \! \approx \! 10^{-5}/\mathrm{K}$.
Hence already a common change $\delta T$ causes a finite drift of roughly 10\,$\mu$m$\cdot(L_{\rm THz}$/1m)$\cdot(\delta T$/1K).
We have chosen a face-to-face geometry with $L_{\rm THz} \! \approx \! 0.2$\,m to keep this term small.
Note that it is not always possible to choose a small value of $L_{\rm THz}$, e.g.,
for measurements with focusing optics or within an optical cryostat.
Additionally, a temperature difference $\Delta T$ between the two arms contributes roughly
10\,$\mu$m$\cdot(L_{\rm Rx}$/1m)$\cdot(\Delta T$/1K). Due to the fiber stretchers, we have to deal with
$L_{\rm Rx}$\,$\approx$\,100\,m, causing a drift of 1000\,$\mu$m$\cdot(\Delta T$/1K).
Therefore it is of utmost importance to prevent a temperature difference between the two arms,
i.e., to provide good thermal contact between the two fiber stretchers.
However, it is not possible to stabilize an extended setup to within 1\,mK, which is already
sufficient to cause a drift of 1\,$\mu$m.

To demonstrate that our normalization method is able to correct for these drifts,
we fix all 3 terahertz frequencies $\nu_i$ with $i \in \{ {\rm ref},31,32\}$
and compare the temporal drifts $\delta\varphi(\nu_i,t)$ measured in a laboratory without temperature stabilization.
Ideally, the phase differences $\Delta \varphi(\nu_i)$ are independent of time.
Accordingly, finite drifts of
\begin{equation}
  \gamma_i(t) = \delta\varphi(\nu_i,t)  \cdot \frac{c}{2\pi \nu_i} \approx \delta L(t)
\label{eq:delL}
\end{equation}
offer a direct view on the temporal drift $\delta L(t)$ of the optical path-length difference,
at least as long as the frequency-dependent contribution $\propto \delta \nu$ is negligible,
see Eq.\ \ref{eq:delPhi}.
Indeed, for $|\Delta L| \! \leq \! 1$\,cm we find that $\gamma_i(t)$ is basically independent
of the frequency $\nu_i$, see Fig.\ \ref{fig:3freqcorr}.
Over roughly 30\,min, we find drifts of $\pm 5\,\mu$m or 40\,$\mu$m in two data sets measured
with (bottom panel) or without (top) good thermal contact between the two fiber stretchers, respectively.
As discussed above, a temperature difference $\Delta T$ between the two stretchers is a major source for
a length drift.
Similar drifts of about 40\,$\mu$m (or 140\,fs$\cdot c$) have been reported for a fiber-based time-domain setup
with 7\,m of fiber in each arm.\cite{Soltani2014}
Achieving a small drift of $\pm 5\,\mu$m requires an optimized setup (small $\Delta T$, $|\Delta L| \! \leq \! 1$\,cm,
$L_{\rm THz} \! \approx \! 0.2$\,m).\cite{Roggenbuck2012}
Plotting $\gamma_i(t)-\gamma_{\rm ref}(t)$ averaged over 10\,s shows that $\delta L(t)$ is determined
to $\pm 1$\,$\mu$m
in case of good thermal contact between the two stretchers (cf.\ lower panel of Fig.\ \ref{fig:3freqcorr}),
even without temperature stabilization of the laboratory.
Monitoring $\delta\varphi(\nu_{\rm ref},t)$ thus allows us to correct the drift $\delta L(t)$ with high accuracy.
The normalized phases
\begin{equation}
  \Delta\varphi_{\rm corr}(\nu_i) =  \Delta\varphi(\nu_i) - \delta \varphi(\nu_{\rm ref}) \cdot \frac{\nu_i}{\nu_{\rm ref}}
\label{eq:corr}
\end{equation}
with $i$\,=\,31,\,32 are nearly insensitive to thermal drifts.

The remaining uncertainty of about 1\,$\mu$m corresponds to $10^{-8} \cdot L_{\rm Rx}$.
This uncertainty is caused by a finite contribution of frequency fluctuations $\delta \nu$,
cf.\ Eq.\ \ref{eq:delPhi}, and by the uncertainty of the interferogram fit.
Moreover, one has to keep in mind that it takes about 300\,ms to measure an interferogram,
thus the method cannot correct fluctuations on such short time scales.
However, the drift correction is very well suited to correct for slower drifts of $\Delta L$
and thus to provide long-term stability as required, e.g., for high-resolution measurements over a broad frequency range
or for measurements as a function of an additional external parameter.

\begin{figure}
\centering
\includegraphics[width=0.85\columnwidth]{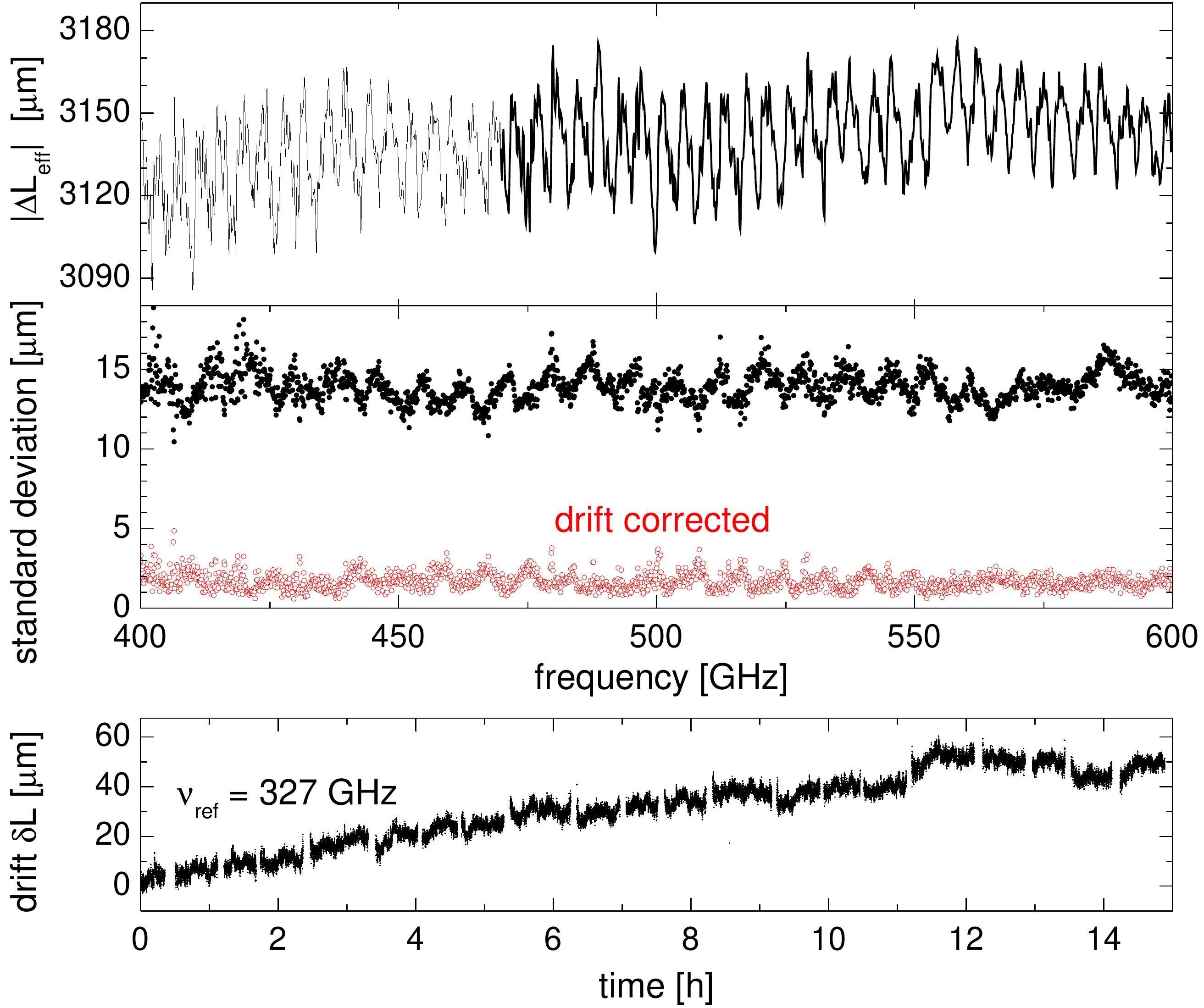}
\caption{Top: Effective optical path-length difference (cf.\ Eq.\ \ref{eq:Leff}) obtained
by averaging over 10 drift-corrected runs measured over 15\,h.
Middle: Standard deviation of the 10 runs with (red) and without (black) drift correction.
Bottom: Drift $\delta L(t)$ determined from $\Delta \varphi$(327\,GHz).
}
\label{fig:sweep}
\end{figure}

As an example, we performed 10 frequency sweeps with a resolution of 100\,MHz over a period of 15\,h without a sample.
The phase data $\Delta \varphi(\nu_{\rm ref})$ measured at the fixed frequency of 327\,GHz reveal
a slow drift $\delta L(t)$ of about 60\,$\mu$m over 12\,h (bottom panel of Fig.\ \ref{fig:sweep}).
The top panel of Fig.\ \ref{fig:sweep} shows the effective optical path-length difference
\begin{equation}
\Delta L_{\rm eff} \, = \, (\Delta \varphi - \Delta \varphi_{\rm an} - \Delta \varphi_{RC}) \cdot c/2\pi \nu \,\,\, ,
\label{eq:Leff}
\end{equation}
where $\Delta \varphi_{\rm an}$ and $\Delta \varphi_{RC}$ denote the contributions stemming
from the group delay of the antennae and the photomixer impedance (see Ref.\ [\onlinecite{Langenbach2014}] for details).
Accordingly, $\Delta L_{\rm eff}$ contains the optical path-length difference $\Delta L \approx -3$\,mm as well as the effects of
water vapor absorption in the terahertz path and standing waves.
The strong modulation of $\Delta L_{\rm eff}$ with a period of 4.07\,GHz is caused by standing waves within the
photomixers' Si lenses.\cite{Langenbach2014}
The data in the top panel of Fig.\ \ref{fig:sweep} were obtained by averaging over 10 frequency sweeps after drift correction,
yielding a standard deviation of $\sigma$\,$\approx$\,1-2\,$\mu$m (red symbols in middle panel).
The accuracy obtained in the frequency sweeps is thus similar to the results for fixed frequencies discussed above.
The modulation period of $\sigma$ of $\sim 8$\,GHz stems from the frequency control of the scanning laser,
i.e., the remaining uncertainty originates mainly from the uncertainty $\delta \nu$.
Averaging the uncorrected data yields a standard deviation of about 15\,$\mu$m (black symbols in middle panel).
This clearly demonstrates the importance of drift correction for an accurate determination of the phase
as well as the applicability of our method to broadband spectroscopy.

In summary, we employ three lasers to perform cw terahertz spectroscopy at one fixed frequency $\nu_{\rm ref}$
and two scanning frequencies $\nu_{31}$ and $\nu_{21}$. The data obtained at $\nu_{\rm ref}$ monitors
the drift $\delta L(t)$ of the optical path-length difference with an accuracy of about 1-2\,$\mu$m.
This can be used to self-normalize the frequency dependence of the phase data measured at the scanning frequencies.
The self-normalization is close to ideal because the drift correction is achieved by comparison of waves
which travelled the same path at the same time, without reduction in measurement speed.
This method allows for a reliable determination of the phase even in situations with large drifts
such as in non-ideal environmental conditions.

\end{document}